\documentclass[onecolumn,amsmath,amssymb]{qm2008_abs}
\usepackage{graphicx}
\usepackage{dcolumn}
\usepackage{bm}
\begin{document}

\title{{STAR measurement of system size and incident energy dependence of $p_{t}$ correlations
at RHIC  }}

\bigskip
\bigskip
\author{\large Lokesh Kumar (for STAR Collaboration)}
\email{lokesh@rcf.rhic.bnl.gov}
\affiliation{Panjab University, Chandigarh, India}
\bigskip
\bigskip

\begin{abstract}
\leftskip1.0cm
\rightskip1.0cm

We present the results on measurement of event-by-event $p_t$ fluctuations and correlations 
for Cu+Cu collisions at $\sqrt{s_{NN}}$  = 62.4 and 200 GeV from STAR experiment at RHIC.
These results are compared with those from Au+Au collisions at $\sqrt{s_{NN}}$  = 62.4 and 200
GeV to study the system size dependence. We find that the dynamical $p_t$
fluctuations are finite and decrease with increasing collision centrality.
The $p_t$  correlations are studied as a function of collision centrality and
 are observed to decrease as we go from peripheral to central collisions. The
square root of $p_t$ correlations scaled by mean $p_t$ are observed to be
independent of beam energy as well as colliding ion size.

\end{abstract}

\maketitle

\section{Introduction}

The study of event-by-event fluctuations in various quantities like average transverse 
momentum, multiplicity, 
and conserved quantities such as net charge, may  provide 
evidence for  formation of the Quark Gluon Plasma (QGP) in relativistic heavy-ion 
collisions~\cite{lok}. The event-by-event transverse momentum fluctuations can be related to the
temperature fluctuations which are  related to specific heat of the system~\cite{lok0}. The
non-monotonic change of $p_{t}$ correlations as a function of centrality and/or as the 
incident energy can also be a possible signal of QGP~\cite{lok}. Several investigations have been
carried out on the study of fluctuations and correlations in relativistic
heavy ion collisions~\cite{lok4}.  Measurements performed by the STAR  collaboration showed that transverse
momentum fluctuations in Au+Au collisions at RHIC energies are
finite but small relative to prediction invoking the formation of
QGP~\cite{lok3}. We present  $p_t$ fluctuations and correlations in Cu+Cu
collisions at 62.4 and 200 GeV and compare with those in Au+Au
collisions at similar energies to investigate the system size
dependence.



\section{Data Analysis }
        Data for this analysis were obtained using the Solenoidal
        Tracker at RHIC(STAR) detector employing a 0.5 T uniform magnetic
        field parallel to the beam axis. We study Cu+Cu collisions at
        $\sqrt{s_{NN}}$ = 62.4 and 200 GeV. The detector used in this analysis
        is the Time Projection Chamber(TPC). The events are required to
        have a vertex within $\pm$ 25 cm along the beam pipe from the center
        of STAR. The charged particle tracks used were required to have
        at least 20 hits in the TPC and have originated within 1 cm of
        the measured event vertex. Events were accepted within 1 cm of
        the center of STAR in the plane perpendicular to the beam
        direction. Tracks from TPC with 0.15 GeV/c $\le$ $p_t$ $\le$ 2.0 GeV/c
        with $\mid$ $\eta$ $\mid$ $<$ 1.0 were used in this analysis. The collision
        centrality is determined by using uncorrected multiplicity of all charged
        particles measured in TPC within $\mid \eta \mid$ $<$ 0.5.
        Each centrality bin was represented by the number of participating
        nucleons, $N_{part}$, using a Glauber Monte Carlo calculations~\cite{lok5}.
         Data analysed are from minimum bias triggers.

         The $p_t$ fluctuations are studied by comparing the mean and rms
        deviations of the event-by event mean $p_t$ ( $<$$p_t$$>$) distributions
        of data and mixed events. Mixed events were constructed by
        randomly selecting one track from an event chosen from the
        measured events in the same centrality keeping the same event
        multiplicity as in the data. We study the dynamical fluctuations
        defined as: $\sigma_{dyn}$ =  $\sqrt{ \sigma_{data}^{2} - \sigma_{mix}^{2}}$. Where 
	$\sigma_{data}$ is the rms value of 
	the $<$ $p_t$ $>$ distribution of data and $\sigma_{mix}$ is the rms value of
	the $<$ $p_t$ $>$ distribution of mixed events.

	The transverse momentum correlations are studied using the following equation~\cite{lok3}:
%
%
\begin{eqnarray}
\langle \Delta p_{t,i} \Delta p_{t,j} \rangle=\frac{1}{N_{\rm event}}
\sum_{k=1}^{N_{\rm event}}{\sum\limits_{j=1,i\ne j}^{N_{k}}}{\frac{{\left( {p_{t,i}-\left\langle {\left\langle {p_t} \right\rangle }
\right\rangle } \right)\left( {p_{t,j}-\left\langle
{\left\langle {p_t} \right\rangle } \right\rangle } \right)} } {N_{k}(N_{k}-1)}}
\end{eqnarray}

where $N_{\rm event}$ is the number of events in a particular centrality, $p_{t,i}$ is the transverse 
momentum of the $i^{th}$ track in each event, $N_k$ is the number of 
tracks in the $k^{th}$ event. The  overall event  average transverse 
momentum $\left\langle {\left\langle{p_t} \right\rangle} \right\rangle$ is given by
$\left\langle {\left\langle {p_t} \right\rangle } \right\rangle =\left( {\sum\limits_{k=1}^{N_{\rm event}} {\left\langle {p_t} \right\rangle _{k}}} \right)/N_{\rm event}$, where $\left\langle {p_t} \right\rangle _k$ is the average 
transverse momentum in the $k^{th}$ event. 

%
%
%

\section{Results and discussion}

\begin{figure}
\includegraphics[scale=0.32]{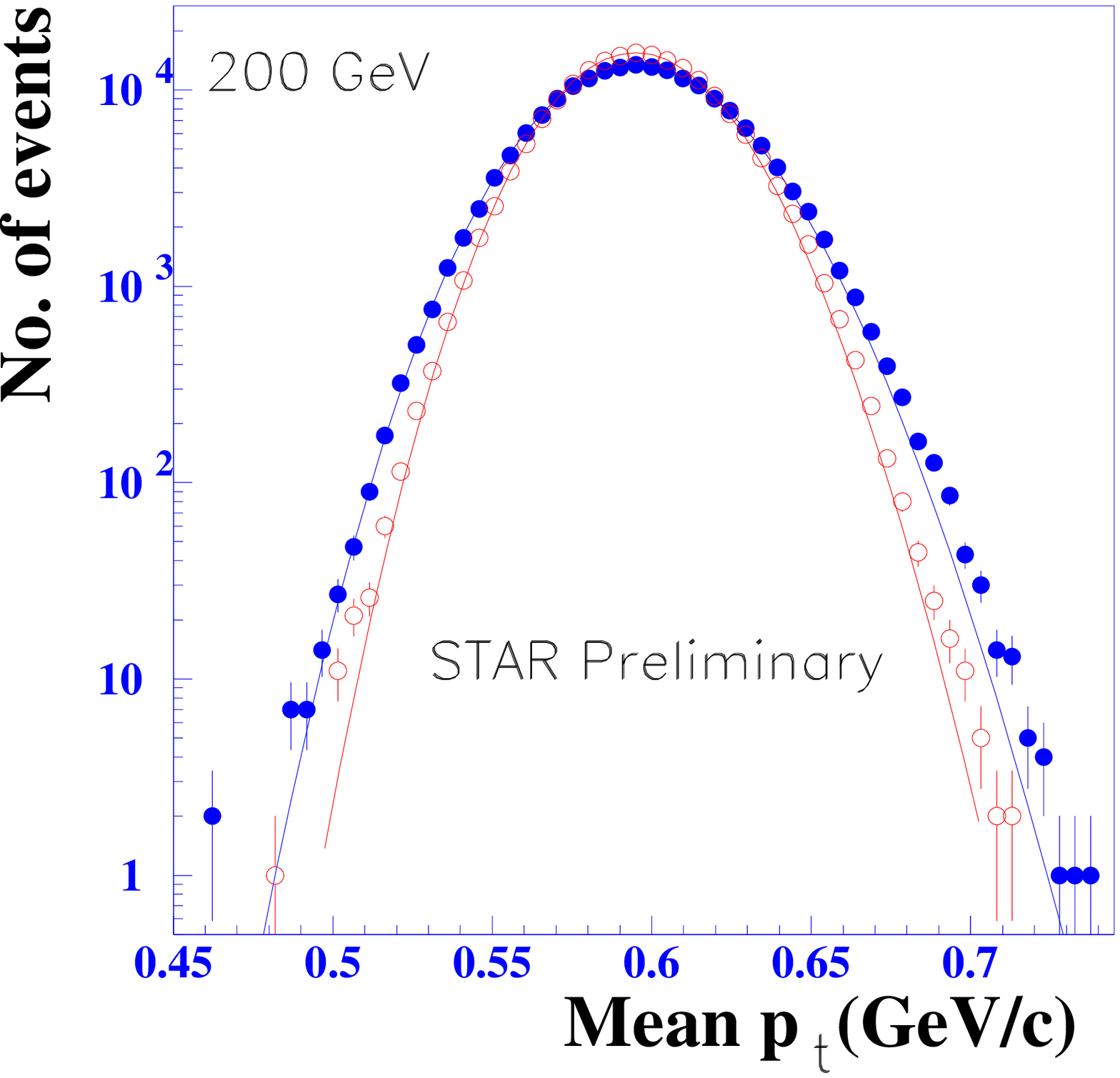}
\includegraphics[scale=0.32]{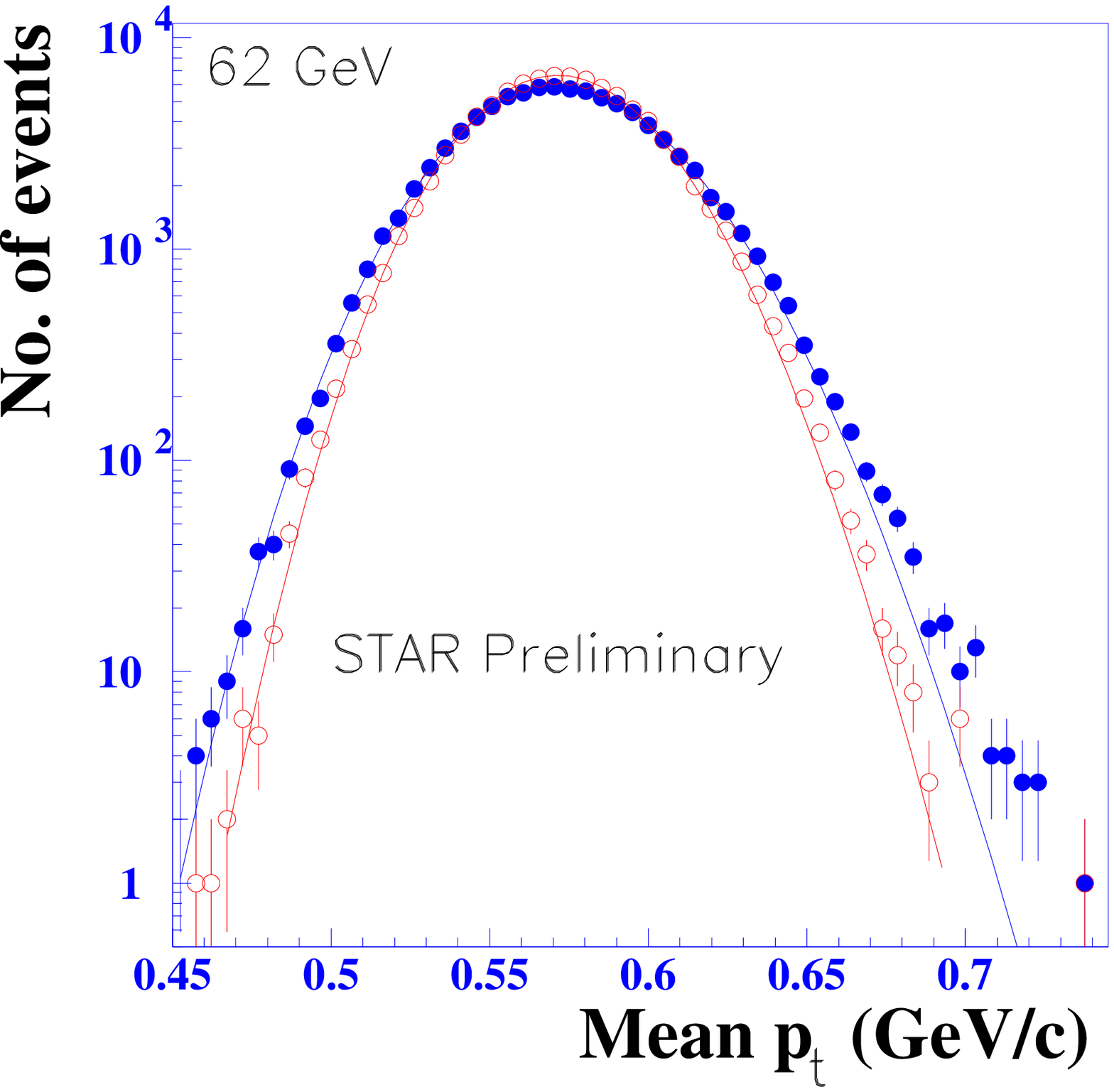}
\vspace{-0.5cm}
\caption{ Mean p$_t$ distributions for data (solid circles) and mixed events
 (open circles) for the 10\% most central events for 
Cu+Cu collisions at 200 GeV (left) and 62.4 GeV(right). The lines represent gamma
distribution.}
\label{fig1}
\end{figure}

The Fig.~\ref{fig1}  shows the event-by-event $\langle p_t \rangle$ 
distributions for data (solid circles)and mixed events (open circles)
for 0-10\% Cu+Cu collisions at 200 GeV (left) and 62.4 GeV (right). The errors
shown in the results are statistical only. Distributions
for the data are wider than those of mixed events indicating the presence
of non-statistical fluctuations at both incident energies. Similar results  are
obtained for other centralities. We find that the dynamical fluctuations 
varies from 4.1\% to 2.3\% as the collision centrality goes from 30-40$\%$
(peripheral) to 0-10$\%$ (most central) at both the energies. The lines in
figures represent gamma distributions given as:

\begin{eqnarray}
f(x)= x^{\alpha-1} e^{-x/\beta}/\Gamma(\alpha) \beta^{\alpha}
\end{eqnarray}
Where $\alpha$ = $\mu^{2}$/$\sigma^{2}$ and $\beta$= $\sigma^{2}$/$\mu$ in
GeV/c. Here $\mu$ is the mean in GeV/c and $\sigma$ is the standard deviation
in GeV/c of the mean $p_{t}$ distribution. Tannenbaum~\cite{lok2} showed that without 
$p_{t}$ cuts the $\alpha$/$<$N$>$ should be 2 following the gamma distribution and
$\beta$$<$N$>$ reflects the temperature of the $p_{t}$ distribution. 
Where $<$N$>$ is the average charged particle multiplicity for a given centrality.
$\alpha$/$<$N$>$
varies from 2.21  to 2.13  at 200 GeV and from  2.37 to 2.27 at 62.4 GeV 
and $\beta$$<$N$>$ varies from  258 MeV/c to 279 MeV/c at 200 GeV and 
from 231 MeV/c  to 252 MeV/c    at
62.4 GeV as the collision centrality goes from 30-40$\%$ to 0-10$\%$. 

\begin{figure}[ht]
\includegraphics[scale=0.35]{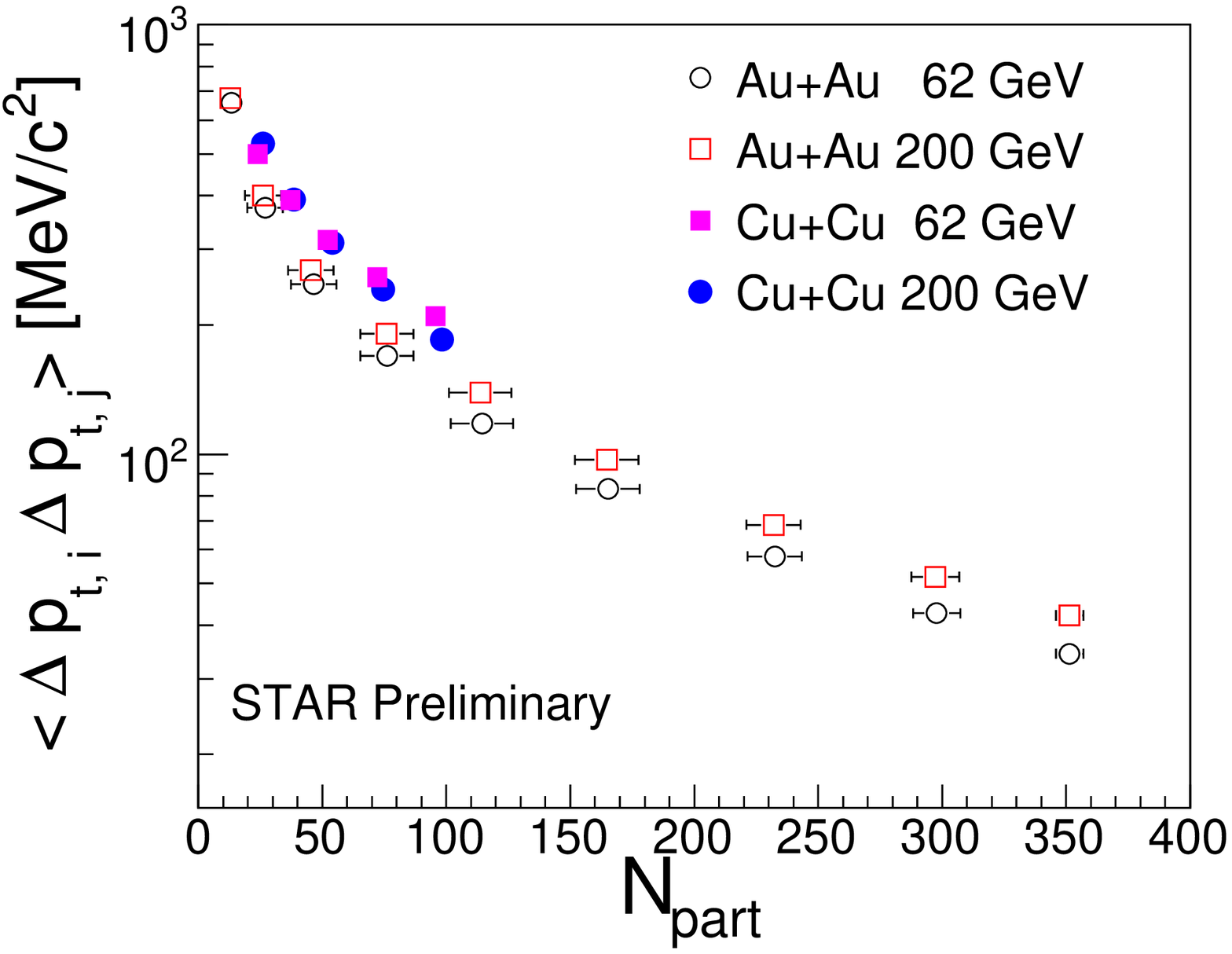}
\includegraphics[scale=0.35]{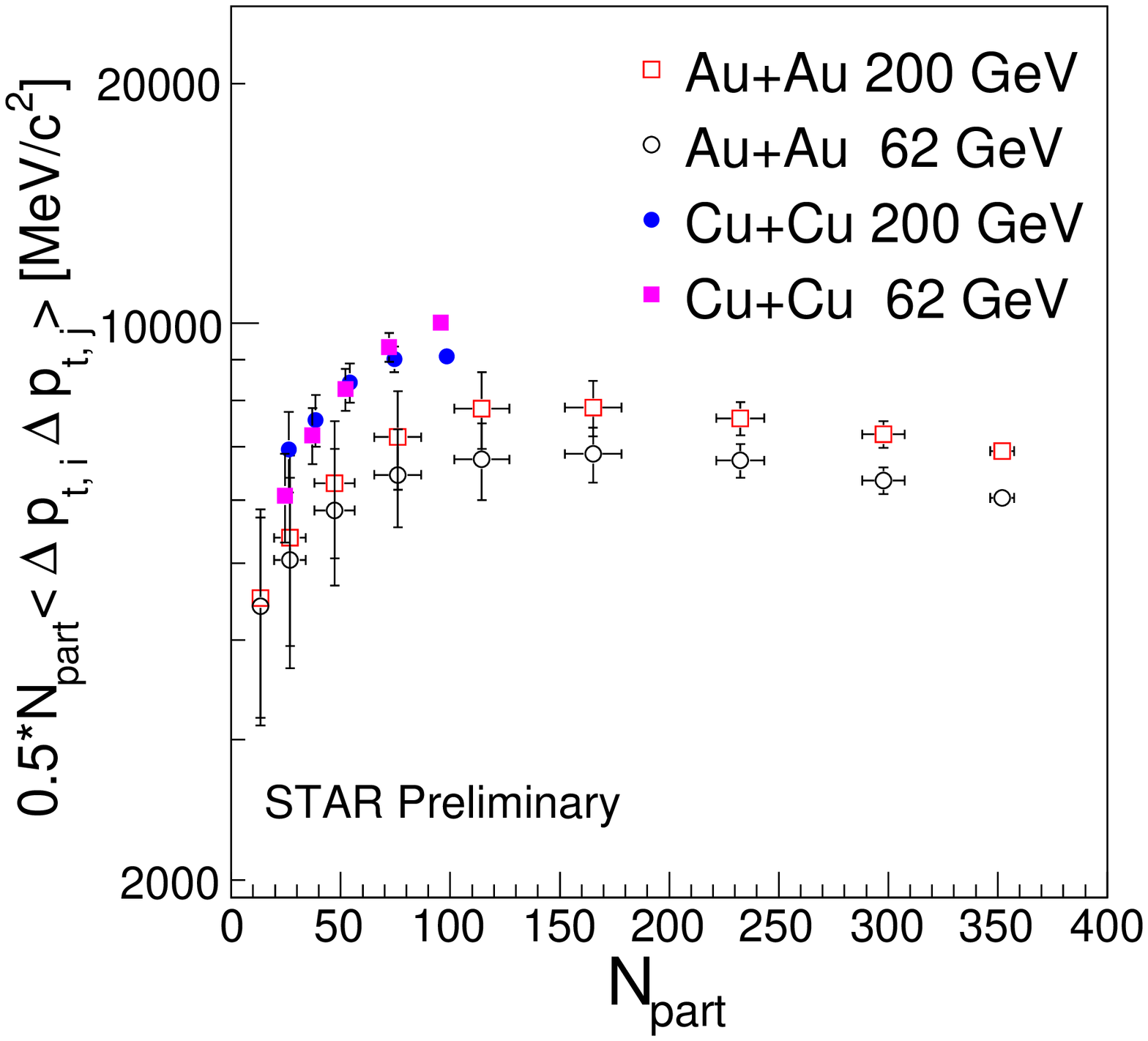}
\vspace{-0.5cm}
\caption{ Left : p$_t$ correlation,  $\langle \Delta p_{t,i} \Delta p_{t,j} \rangle$, as 
a function of $N_{part}$ for Au+Au~\cite{lok3} and Cu+Cu collisions at  
$\sqrt{s_{NN}}$ = 62.4 and 200 GeV.
 Right : p$_t$ correlation mulitplied by half $N_{part}$ as a function of $N_{part}$ for 
Cu+Cu and Au+Au collisions at $\sqrt{s_{NN}}$ = 62.4 and 200 GeV. }
\label{fig2}
\end{figure}

 The Fig.~\ref{fig2} (left) shows the two particle transverse momentum correlations 
obtained using eqn.1   as a function of $N_{part}$  for Cu+Cu at 62.4 GeV, Cu+Cu at 
200 GeV, Au+Au at 62.4 GeV and Au+Au at 200 GeV  collisions~\cite{lok3}. 
The $p_t$ correlations are finite and decrease with increase in number of  participants 
for  both Cu+Cu and Au+Au collisions. This decrease in $p_t$ correlations could be
 due to  correlations being dominated from pair of particles coming from the
 same nucleon-nucleon collision which get diluted with increasing the number of participants.
 The  correlation seems to grow with decreasing system size having similar number of 
participants.
 Fig.~\ref{fig2} (right) displays measured correlation $\langle \Delta p_{t,i} \Delta p_{t,j} 
\rangle$ 
scaled by the number of participating nucleons divided by 2 to
account for the dependence of the correlation on the number of sources of
particle emission. It is seen that the correlation increases with increasing collision
centrality and then saturate in central Au+Au collisions indicating sign of
thermalization~\cite{lok1} whereas in Cu+Cu collisions correlation does not seems to
saturate with increasing collision centrality as system size is much smaller 
as compared to Au+Au. 

\begin{figure}
\includegraphics[scale=0.35]{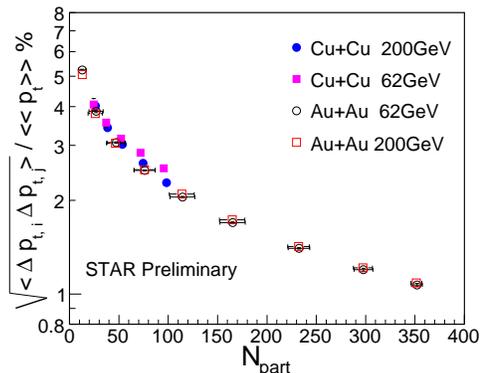}
\vspace{-0.5cm}
\caption{$\sqrt{ \langle \Delta p_{t,i} \Delta p_{t,j} \rangle}/ << p_t >>$ versus $N_{part}$ for 
Au+Au\cite{lok3} and Cu+Cu collisions at $\sqrt{s_{NN}}$ = 62.4 and 200 GeV.}
\label{fig3}
\end{figure}

  The observed dependence of correlation on energy, system size or centrality
may be due to the changes in $<$ $p_{t}$ $>$ with beam energy,centrality and 
colliding ion species. Therefore, we study the square root of the correlation
scaled by the $p_t$ as a function of number of participants (Fig.~\ref{fig3}). It is seen
that the quantity $\sqrt{ \langle \Delta p_{t,i} \Delta p_{t,j} \rangle}/ << p_t >>$ is 
independent of beam energy as well as system size but decreases with
increasing $N_{part}$.

\section{Summary}

We find non-statistical fluctuations in the mean $p_{t}$ at both the
energies. The dynamical fluctuations decrease with increase in collision
centrality. The mean $p_{t}$ distributions follow gamma distributions. The
$p_{t}$ correlations are finite and decrease with increasing in the
number of participants in Cu + Cu collisions at $\sqrt{s_{NN}}$ = 62.4 and 
200 GeV  as was observed in Au + Au collisions. The $p_{t}$ correlation
scaled by the number of participating nucleons in Au+Au collisions saturates
 at higher collision centrality indicating the sign of thermalization~\cite{lok1} whereas
no such saturation is observed in case of Cu + Cu collisions. The quantity
 $\sqrt{ \langle \Delta p_{t,i} \Delta p_{t,j} \rangle}/ << p_t >>$ seems to
be independent beam energy as well as system size.

\end{document}